\documentclass[,journal]{IEEEtran}
\usepackage{amsmath,amsfonts}
\usepackage{algorithmic}
\usepackage{algorithm}
\usepackage{array}
\usepackage[caption=false,font=normalsize,labelfont=sf,textfont=sf]{subfig}
\usepackage{textcomp}
\usepackage{stfloats}
\usepackage{url}
\usepackage{verbatim}
\usepackage{graphicx}
\usepackage{cite}

\usepackage{amsmath}
\usepackage{multirow}
\usepackage[table,xcdraw]{xcolor}
\usepackage{booktabs}

\newcommand{\sunnycyc}[1]{{\color{black}#1}}

\newcommand{\sunny}[1]{{\color{black}#1}}
\newcommand{\cyc}[1]{{\color{black}#1}}
\newcommand{\cy}[1]{{\color{black}#1}}

\begin{document}
\title{Local \sunny{Periodicity}-Based Beat Tracking for Expressive Classical Piano Music}

\author{Ching-Yu Chiu, Meinard M\"uller,  Matthew E. P. Davies,  Alvin Wen-Yu Su, and Yi-Hsuan Yang

\thanks{Meinard M\"uller is supported by the International Audio Laboratories Erlangen \sunnycyc{--} a joint institution
of the Friedrich-Alexander-Universität Erlangen-Nürnberg (FAU) and Fraunhofer Institute for Integrated Circuits IIS. Matthew E. P. Davies was supported by FCT--Foundation for Science and Technology, I.P., through the Project MERGE through the National Funds (PIDDAC) through the Portuguese State Budget under Grant PTDC/CCI-COM/3171/2021; and in part by the European Social Fund through the Regional Operational Program Centro 2020 Project CISUC under Grant UID/CEC/00326/2020.
}
\thanks{Ching-Yu Chiu is with the Graduate Program of Multimedia Systems and Intelligent Computing, National Cheng Kung University and Academia Sinica, Taiwan (e-mail: sunnycyc@citi.sinica.edu.tw).}
\thanks{Meinard M\"uller is with the International Audio Laboratories Erlangen, Germany (e-mail: meinard.mueller@audiolabs-erlangen.de).}
\thanks{Matthew E. P. Davies is with the Department of Informatics Engineering, Centre for Informatics and Systems of the University of Coimbra, University of Coimbra, Portugal (e-mail: mepdavies@dei.uc.pt).}
\thanks{Alvin Wen-Yu Su, is with the Department of Computer Science and Information Engineering, National Cheng Kung University, Taiwan (e-mail: alvinsu@mail.ncku.edu.tw).}
\thanks{Yi-Hsuan Yang, is with Yating Music Team, Taiwan AI Labs, Taiwan. He is also with Research Center for IT Innovation, Academia Sinica, Taiwan (e-mail: yang@citi.sinica.edu.tw).}}

\markboth{Journal of \LaTeX\ Class Files,~Vol.~14, No.~8, April~2022}%
{Shell \MakeLowercase{\textit{et al.}}: A Sample Article Using IEEEtran.cls for IEEE Journals}


\maketitle

\begin{abstract}
To model the periodicity of beats, state-of-the-art beat tracking systems use ``post-processing trackers'' (PPTs) that 
rely on several empirically determined global \sunnycyc{assumptions} for tempo transition,
which work well for music with a steady tempo. 
For expressive classical music, however, these assumptions can be too rigid.
With two large datasets of Western classical piano music, namely the Aligned Scores and Performances (ASAP) dataset and a dataset of Chopin's Mazurkas (Maz-5), we report on experiments showing the failure of existing PPTs to cope with local tempo changes, thus calling for new methods. 
\sunny{In this paper, we propose a new local periodicity-based PPT}, called predominant local pulse-based dynamic programming (PLPDP) tracking, that allows for  more flexible tempo \sunnycyc{transitions}. 
Specifically, the new PPT incorporates a method called ``predominant local pulses'' (PLP) \sunnycyc{in combination with} a dynamic programming (DP) \sunnycyc{component} to jointly consider the locally detected periodicity and beat activation strength at each time instant. 
Accordingly, PLPDP accounts for the local periodicity, rather than
\sunnycyc{relying on a global tempo assumption}.
\sunnycyc{Compared to existing PPTs, PLPDP particularly enhances the recall values at the cost of a lower precision, resulting in an overall improvement of F1-score for beat tracking in ASAP (from 0.473 to 0.493) and Maz-5 (from 0.595 to 0.838).}
\end{abstract}



\section{Introduction}
\label{sec:intro}

\IEEEPARstart{B}{eats}, which are generally referred to as a sequence of perceived pulses at the temporal level that a listener would tap to, are fundamental to the understanding of music \cite{DP2007, Dixon2000}. The ability to perceive beats not only allows us to follow the music, but also serves as the basis \sunnycyc{for decomposing, reconstructing, or interacting} with music.
Computationally,
a variety of downstream applications are related to, and could be enhanced by beat tracking  \cite{benetos19spm, huang20mm}. 

Existing beat tracking systems are mainly composed of two parts. \sunnycyc{First, a \emph{novelty detection} module generates a so-called ``novelty function'' (sometimes also called ``activation function''), which is}
a continuous-valued curve that captures the energy or spectral changes over time so as to reveal beat candidates. \sunnycyc{Second,} a \emph{post-processing tracker} (PPT) gives the final binary decision regarding beat occurrence \cite{Fuentes2018, Bock2016d, Bock2020}. While traditional model-based systems usually derive the novelty function based on onset detection methods \cite{Grosche2011, bello2005, zhou2008, klapuri1999, Klapuri2006}, more recent deep learning (DL)-based systems use a \sunnycyc{feature-learning network} to compute directly from audio signals \sunnycyc{an activation function,} indicating the likelihood of observing a beat at each time instant.
\sunnycyc{Due} to the fact that existing novelty detection methods do not handle well the periodic nature of beat times \cite{oyama2021}, existing beat tracking systems generally rely on periodicity-aware PPTs to determine the beat positions. 
Motivated by the design of \cite{Bock2016d, Krebs2015},
most existing PPTs are state-space models (see Section \ref{sec:related:ppt} for more details), such as dynamic Bayesian network (DBN), hidden Markov model (HMM) \cite{Krebs2015, Bock2016d, oyama2021}, conditional random field (CRF) \cite{Magdalena2019}, or particle filtering
\cite{heydari2021}.

DL-based beat tracking systems have achieved great success for music with steady tempo, in particular for pop, rock, and dance music \cite{Bock2016d, Bock2016mm, Bock2020}. However, probably due to the scarcity of publicly available data, the performance of state-of-the-art DL-based beat trackers for expressive classical music has seldomly been discussed, or is far from satisfactory if reported.  Specifically, the performance of state-of-the-art  DL-based systems for beat or downbeat tracking tends to be 20--30\% worse for classical music than for other music genres \cite{spl21chiu, eusipco21chiu,Durand2016DownbeatDW}. In this paper, we are interested in finding out the underlying reasons for the poor performance while improving beat tracking for expressive classical music.

\sunnycyc{The challenges of beat tracking for expressive classical music are closely related to the properties of the novelty function\cite{Grosche2010b, Holzapfel2012}. For example, the novelty function may get aperiodic due to substantial tempo, rhythmic, or note density variations. Its intensity may get weaker due to blurred note onsets and soft note transitions for non-percussive instruments such as violins or singing. It is also assumed that one may improve the accuracy of beat tracking for classical music by using a larger classical music training set\cite{spl21chiu, Schreiber2020a}. As one contribution of this paper, we present experiments on two large datasets of expressive classical piano music, the Chopin Mazurkas (Maz-5) \cite{Grosche2010b} and the Aligned Scores and Performances (ASAP) dataset \cite{asap-dataset}, and show that there are fundamental issues related to the PPT that needs to be addressed first.}


\begin{figure*}
\centering
	\includegraphics[width = 1.4\columnwidth]{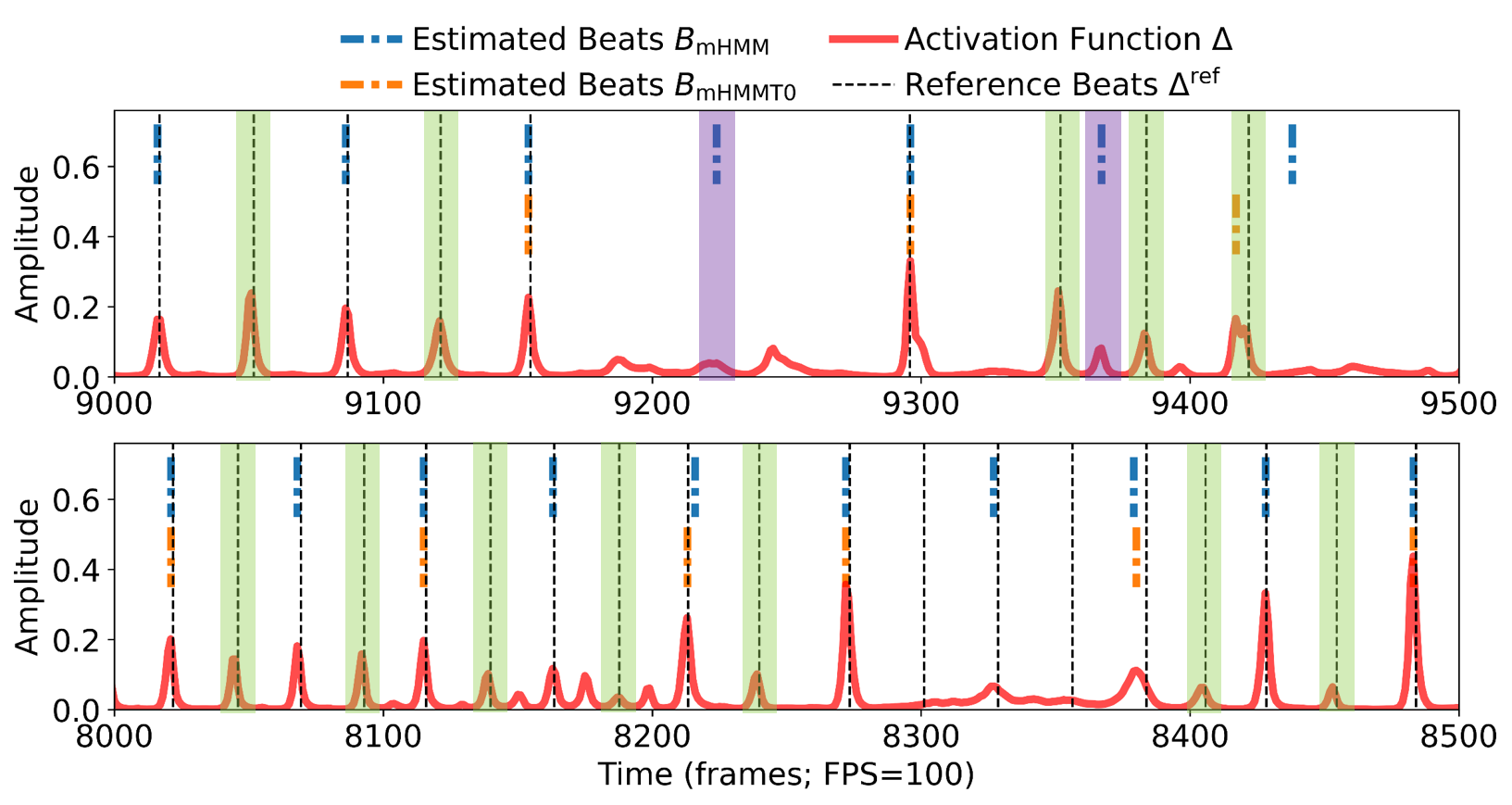}
	\caption{
 \sunnycyc{Beat tracking result of a state-of-the-art DL-based system from \texttt{madmom} for two recordings of classical music, one from Maz-5 \cite{Grosche2010b} \textbf{(top)} and the other from ASAP \cite{asap-dataset} \textbf{(bottom)}. The activation function (using a frame rate of FPS $=100$) is shown in red. Green shades highlight the false-negative estimations with activation peaks. Purple shades highlight the false-positive estimations caused by non-beat activation peaks. Best viewed in color.}
	}
	\label{fig:teaser}
\end{figure*} 

Figure \ref{fig:teaser} illustrates the result of beat tracking for two recordings of classical music, one from Maz-5 and the other from ASAP,
using the system from the \texttt{madmom} library \cite{Bock2016d, Bock2016mm}.
The system uses an ensemble of recurrent neural networks (RNN) for estimating the beat activation functions,
and an HMM-based PPT \cite{Krebs2015} for ensuring the periodicity of the detected beats. \cy{As mentioned, existing PPTs for beat tracking 
\sunnycyc{closely follow} the HMM-based PPT in \texttt{madmom} library \cite{Bock2016d, Krebs2015}. As such, we adopt this PPT as one of our main baseline models, using the default (and widely-used) parameter setting of \texttt{madmom}. We refer to this PPT as ``mHMM'' hereafter. Besides, in our study we also consider a more flexible version of mHMM where we tune its parameter setting  to allow for \sunnycyc{a} higher tempo transition \sunnycyc{likelihood}, and refer to it as ``mHMMT0'' (see Sections \ref{sec:ppt_settings} and \ref{sec:real_exp} for details).}
We can see from the reference beats (i.e., the ground truth annotations) that the two recordings feature different degrees of tempo variation. 
We also see that the system has many false positives and false negatives, caused not only by its imperfect beat activation function (e.g., activation peaks at non-beat positions), but also by the HMM-based PPT.
For both recordings, \cy{mHMM} \sunny{assumes} that there is a relatively stable (and slower) tempo and chooses to ignore local activation peaks corresponding to true beat positions. Similar \sunnycyc{detection errors} can be observed from the result of mHMMT0.
Moreover, our analysis (see Section \ref{sec:synth_exp}) shows that, even with a \emph{perfect} oracle activation function created synthetically with the reference beats, the average F1 score obtained by the \sunnycyc{mHMM-based estimations} across the 301 recordings of Maz-5 remains lower than 0.80. 
This suggests that the PPT makes some tempo assumptions that do not work for expressive music. 

In cognitive neuroscience, it has been found that the human brain generally focuses more on local events 
\cy{(e.g., musical onsets within a small time window)} 
and continuously tries to predict incoming information \cite{clark_2013}. Given a few musical onset events, we may start having expectations for the incoming events \cite{Bouwer2020, Obleser2017, Nobre2017}. These expectations are to be adjusted, strengthened, or abandoned based on the consistency between the expectations and the incoming events. These ``temporal expectations'' may be part of the reasons why humans can,  to a certain degree, adaptively deal with tempo variations, syncopation, and rest notes on beat positions in music. 

In light of the above observations, 
we develop a new PPT method that 
\sunny{mimics}
the temporal expectations computed from the beat activation function.
Specifically, we propose to use a feature called ``predominant local pulses'' (PLP) \cite{Grosche2011, fmp2021, meier2021realtime} (see Section \ref{sec:PLPDP} for details) to estimate information regarding local 
\sunny{periodicity}
(analogous to temporal expectations) from the beat activation function. 
The PLP curve is converted into two curves, one of which contains information of local inter-pulse-intervals (IPIs) and the other containing the confidence of locally detected periodicity.
Then, we propose a new dynamic programming (DP)-based PPT that takes the two curves as time-varying tempo conditions to track the beats. 
With quantitative experiments, using both real and synthetic activation functions on Maz-5 and ASAP (Section \ref{sec:results}), we demonstrate the \sunnycyc{advantages} of the new PPT method, named ``PLPDP,'' over representative existing PPTs for beat tracking of expressive music with continuously varying tempo.
We also identify some  limitations of PLPDP that need to be further addressed in future work (Section \ref{sec:conclusions})\footnote{\sunnycyc{For reproducibility, we provide open-source code for PLPDP at: \url{https://github.com/SunnyCYC/plpdp4beat/}. We also have a project web page that provides examples of the beat tracking results: \url{https://sunnycyc.github.io/plpdp4beat-demo/}.}}.

\section{Related Works}
\label{sec:related_works}
\subsection{Beat and Tempo for Expressive Classical Music}
Research on beat tracking for classical music with large tempo variations dates back two decades
\cite{Dixon2000, Dixon2001}. However, it was not until the work by Grosche \emph{et al.}  \cite{Grosche2010b, Grosche2011} that larger evaluations 
were carried out.
In their first publication \cite{Grosche2010b}, they discussed five musical properties that cause problems for beat tracking, and subsequently conducted \sunny{systematic} experiments to analyze and make the limitations of state-of-the-art beat trackers explicit. As beat tracking systems at that time relied more on the quality of \sunnycyc{the underlying} novelty function, the influence of different musical properties on the novelty detection was illustrated. However, \sunnycyc{the influence of musical properties on the performance of} PPT was not explicitly investigated.
In the publication \cite{Grosche2011}, the idea of predominant local pulse (PLP) was introduced and used for modeling the local periodicity of model-based onset novelty function. \sunnycyc{Specifically, PLP extracts and enhances the local periodicity of the input novelty function via analyzing the onset peaks within small windows, which motivates the core idea of this paper (see Section \ref{sec:PLPDP}).}
\cyc{The PLP-enhanced novelty function can be combined with a PPT based on dynamic programming (DP), \sunnycyc{e.g., as} introduced by Ellis \cite{DP2007}. However, assuming an overall constant tempo (see Section \ref{sec:related:ppt}), the DP-based PPT cannot handle strong tempo variations.}

More recently,
motivated by the work of B{\"{o}}ck \emph{et al.} 
\cite{Bock2016d, Bock2016mm},
a variety of DL-based feature learning networks have been proposed \cite{Davies2019, Bock2020}.
Moreover, as DL models typically require large training data,
only a few studies tackle beat tracking-related tasks for classical music \sunnycyc{where beat annotations are hardly available}.
For example, 
Schreiber \emph{et al.} \cite{Schreiber2020a} pioneered the use of DL-based methods for 
modeling local tempo of Maz-5.
Specifically, they aggregated several beats into a higher-level local tempo representation, and used that local tempo representation as the target of their DL-based model.
\sunnycyc{In other words,} their work \sunnycyc{aims at estimating the} local tempo, \sunnycyc{but} not for predicting the individual beats.

\subsection{\cy{PPTs with Global Assumptions for Tempo}} 
\label{sec:related:ppt}
\sunnycyc{The dynamic programming-based beat tracker (DP) as introduced by Ellis\cite{DP2007}} is a widely-used PPT that we consider as a baseline in our evaluation. The DP method assumes that the musical piece is performed with a roughly constant tempo and that the activation is high at beat positions. DP introduces a score that jointly considers how well an input \sunnycyc{novelty} function fits the two assumptions and finds \sunnycyc{globally} the best beat sequence that maximizes the score via a \sunnycyc{dynamic programming} algorithm. \sunnycyc{Based on} the constant tempo assumption, a global tempo value is \sunnycyc{used to balance out}
the consistency between target and estimated inter-beat-interval (IBI) (see Section \ref{sec:plp_dp_comb} for details). The global tempo value can be derived either from the mean IBI of the reference beats \cite{Grosche2011}, or via auotocorrelation-based tempo estimation methods 
\cite{DP2007, context_dep2007}. 

While the constant tempo assumption grants DP a simple formulation and realization, it also limits \cyc{the PPT's} flexibility. Aiming 
\sunnycyc{at}
jointly modeling \sunnycyc{tempi} and beats, Krebs \emph{et al.} \cite{Krebs2015} 
extended
the ``bar pointer model'' \cite{Whiteley2006, Whiteley2007}, 
and proposed a refined state-space \sunny{discretization} and tempo transition model. Their main contributions are the design of \sunnycyc{a} state-space discretization \sunnycyc{model} that ensures sufficient tempo resolution for each hidden state and time resolution consistency between hidden states \sunnycyc{of different tempi}, and a new transition model \sunnycyc{based on a first-order Markov assumption to} improve the \sunnycyc{stability of} tempo trajectories. 
\sunnycyc{In particular, they} increased the tempo stability by proposing a transition model that only allows tempo transitions at beat positions and empirically adopted an exponential distribution function as the tempo transition \sunnycyc{likelihood} function (see Section \ref{sec:HMMtrans_model} for details).

Due to their success in both reducing computational cost and outperforming the original model, existing mainstream PPTs \cite{Bock2016d, heydari2021, heydari2021dlb, kazuhiko_yamamoto_2021, oyama2021, Magdalena2019} are mainly motivated by \cite{Krebs2015}. These PPTs may differ in their optimization mechanisms \cite{heydari2021dlb} or the use of extra information (e.g., beat phase \cite{oyama2021} or time signature \cite{Bock2016d}), but they generally adopt similar empirically determined \sunnycyc{tempo transition likelihood functions based on a first-order Markov assumption.}


We note that in related studies on rhythm transcription, \sunnycyc{researchers also apply} HMM-based methods to take an input signal and estimate the metrical positions of all musical notes. In particular, similar first-order Markov assumptions and tempo transition probability distributions are adopted \cite{shibata2021, Eita2018, Eita2014}. Despite that local tempo and local tempo changes are parameterized in \sunnycyc{these} models, parameters are determined globally (e.g., based on a dataset) without the knowledge of ``local periodicity" (which is explicitly extracted based on small local windows) proposed in this work. \sunnycyc{Focusing} on beat tracking for expressive classical piano music, we limit our discussion of HMMs to those proposed for beat tracking \sunnycyc{and not for the considered HMMs for rhythm transcription.}

\begin{figure}
\centering
	\includegraphics[width = 0.9\columnwidth]{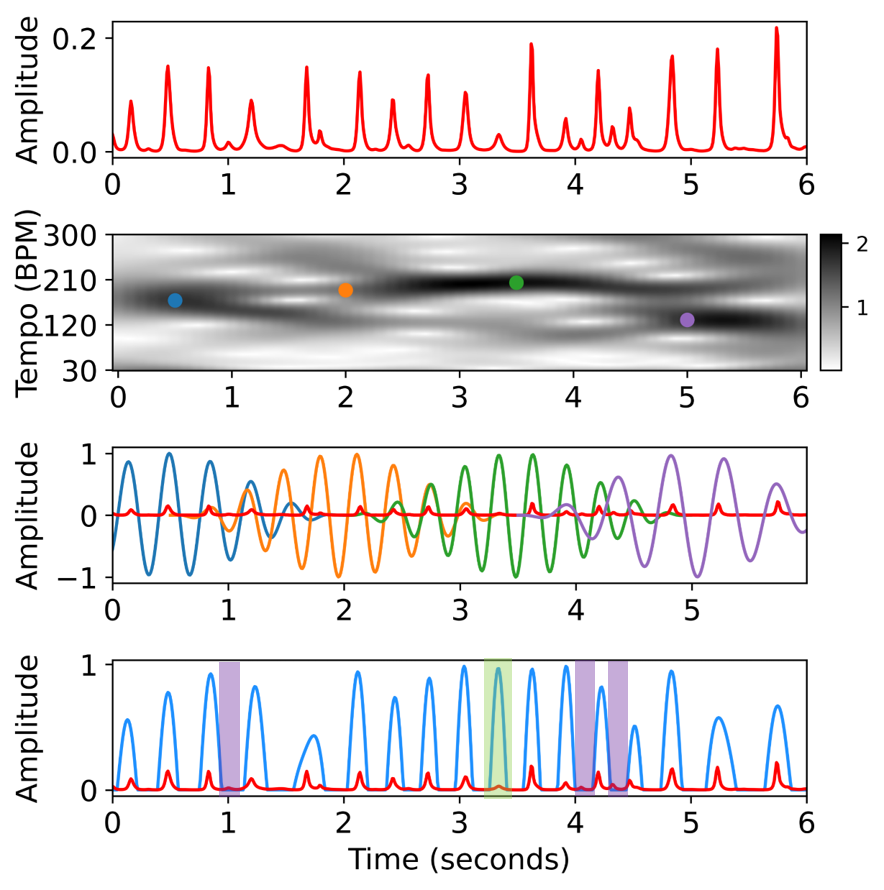}
	\caption{
	Illustration of the workflow of \sunnycyc{the PLP calculation}. \textbf{(a)} The novelty function (red curve\sunnycyc{; also plotted in (c) and (d) with smaller amplitude for alignment comparison}) computed from a recording. \textbf{(b)} Fourier tempogram \sunny{with four colored dots indicating the corresponding time positions of the \sunnycyc{optimal} sinusoidal kernels in (c)}. \textbf{(c)} Optimal sinusoidal kernels at four different time positions. \textbf{(d)} PLP (blue curve) derived via overlap-adding and half-wave rectification. \sunnycyc{Purple shades highlight the novelty peaks suppressed by PLP. Green shade highlights the novelty peak enhanced by PLP.}
	}
	\label{fig:plp_ex}
\end{figure}

\section{\cyc{PLPDP-Based PPT}}
\label{sec:PLPDP}

In this section, we provide the details of the proposed PLPDP-based PPT. We first introduce \sunnycyc{the PLP concept} \cite{Grosche2011, fmp2021, meier2021realtime} and then elaborate \sunnycyc{on} the similarity between PLP and human temporal expectations.
Next, we describe a method to reduce the artifacts of PLP curves at regions with tempo variations.
Subsequently, we demonstrate how to derive the tempo-related information \cyc{reflecting} local temporal expectation and confidence from a PLP \sunnycyc{function}. Finally, we present the algorithm that connects DP with the two tempo-related conditions to realize the PLPDP-based PPT. 

\subsection{PLP as Local Temporal Expectations}
\label{sec:plp} 
As mentioned in Section \ref{sec:intro}, we found that the PLP, a method originally proposed to model and enhance the periodicity of \sunny{musical onset novelty functions} \cite{Grosche2011},
shows behaviors that are interestingly akin to human temporal expectations \sunnycyc{\cite{Xu2020}}. 
Figure \ref{fig:plp_ex} demonstrates the computation and the ``temporal expectations'' of PLP given \sunnycyc{a} pre-computed novelty function.
The main idea of PLP is to generate periodic pulses that align with a given novelty curve based on a local periodicity assumption. 
This is achieved by locally comparing the novelty curve with windowed sinusoidal kernels, and accumulating \sunnycyc{over time} all the optimal sinusoidal kernels that best capture the local peak structure of the novelty function. 
Specifically,
the computation of PLP begins with computing the ``Fourier tempogram'' \cite{tempogram} (cf. Figure \ref{fig:plp_ex}b) via a short-time Fourier transform \sunnycyc{(STFT)}, for a certain pre-determined tempo range (e.g., $\theta \in [30:300]$ beats-per-minute; BPM).
Then, given pre-determined values of \sunnycyc{the STFT parameters,} kernel size $\kappa$ and hop size $h$, the predominant tempo and phase information of the optimal sinusoidal kernel of each time instant can be derived from the tempogram and the underlying \sunnycyc{complex-valued} Fourier coefficients \cite{fmpplp}. \sunnycyc{For example,} Figure \ref{fig:plp_ex}c shows the optimal sinusoidal kernels at four time positions with $\kappa=3$ (seconds). Finally, via overlap-adding and half-wave rectification (keeping only positive parts of the curve), the PLP curve can be derived.
Note that the peaks of PLP somehow indicate PLP's ``expectation'' of the existence of novelty peaks (e.g., see Figure \ref{fig:plp_ex}d how \sunnycyc{the PLP peaks align with the peaks of the novelty function).} 
Even if some novelty peaks are \sunnycyc{low}, as long as the locally detected periodicity has high confidence, PLP still generates strong peaks (see the peak underlined with green shade in Figure \ref{fig:plp_ex}d). This is similar to how humans can still tap on \sunnycyc{weak or even} rest notes based on temporal expectations. \sunnycyc{Furthermore}, novelty peaks that do not match the detected local periodicity \sunnycyc{are} suppressed (see the regions indicated by purple shades). Moreover, as exemplified in Figure \ref{fig:plp_ex}d, PLP \sunnycyc{has lower peaks (being less confident)}
when the neighboring musical events are not consistent in periodicity. 
\sunny{Inspired by the human ability to adaptively adjust expectations and confidence concerning current and future events, we incorporate the idea of PLP in our beat trackers to mimic this ability.}

\subsection{PLP Curves and Combination}
\label{sec:plp_integration}

\begin{figure}[t]
\centering
	\includegraphics[width = 0.9\columnwidth]{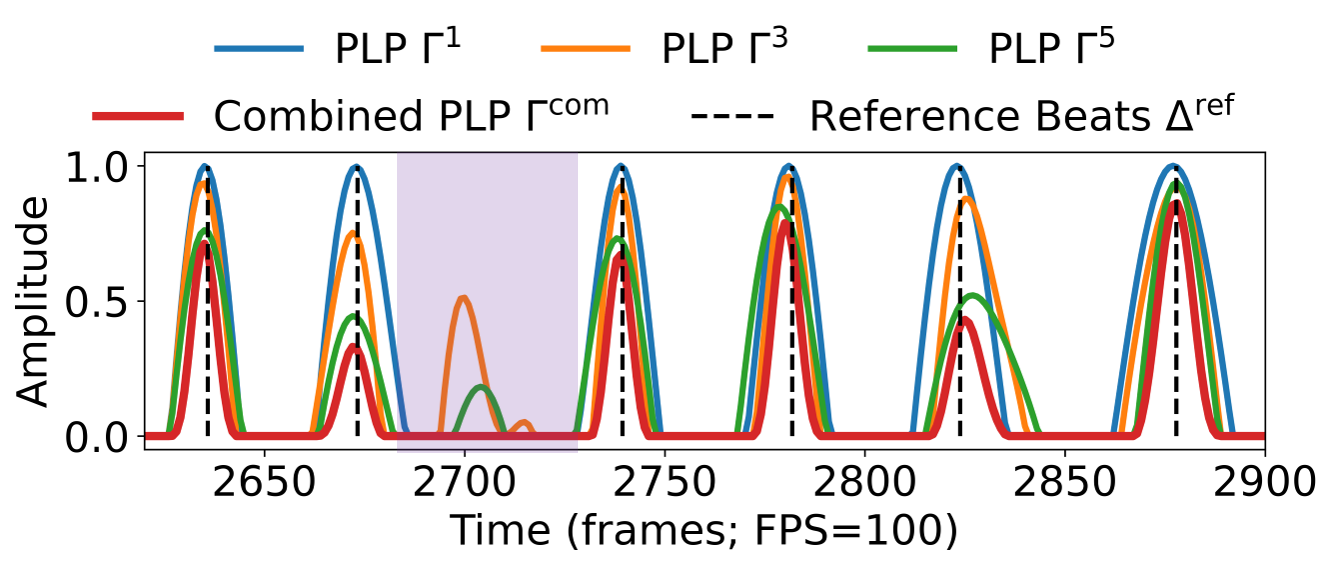}
	\caption{
 \sunnycyc{PLP functions with different kernel sizes $\kappa=1$ (blue), $\kappa=3$ (orange), and $\kappa=5$ (green), and the combined PLP function (red), computed using the oracle novelty function created from the reference beats $\Delta^{\mathrm{ref}}$ (dashed vertical lines). Purple shade highlights the region with a larger tempo/IBI variation where extra peaks of \sunnycyc{individual} PLP functions are suppressed by \sunnycyc{the} combined PLP function. Best viewed in color.}
 }
	\label{fig:plp_cur}
\end{figure} 

The ``\cyc{local sensitivity}'' of a PLP \sunnycyc{function} is largely determined by the choice of the kernel size $\kappa$. 
\sunnycyc{Let $\Gamma^{\kappa}:[1:N]\rightarrow [0, 1]$} 
denote \sunnycyc{the} PLP \sunnycyc{ function for kernel size} $\kappa$ (given in seconds), where $[1:N]:=\{1, 2, ..., N\}$, $N\in\mathbb{N}$, represents the sampled time axis with respect to a fixed sampling rate. \sunnycyc{In our experiments, we use a rate of 100 frames per second.}
\sunnycyc{As an example, Figure \ref{fig:plp_cur} shows $\Gamma^{\kappa}$ for three different kernel sizes $\kappa \in \{1, 3, 5\}$},
using the oracle novelty function 
\sunnycyc{$\Delta^{\text{ref}}:[1:N]\rightarrow [0,1]$}
created from its reference beats as input.\footnote{PLP can take as input either a real or synthetic novelty function (cf. Sections \ref{sec:real_exp} and \ref{sec:synth_exp}). We use a synthetic one here (created from reference beats; see Section \ref{sec:synth_exp} for details) to show the sensitivity of PLP to tempo variations.} It can be seen that at regions with relatively stable tempo, PLP curves \sunnycyc{with different kernel sizes} generally have similar ``temporal expectations'' for peak positions, though with different confidences (i.e., peak heights). However, for regions with a larger tempo variation, \sunnycyc{the three PLP functions show incosistent behaviors}
(e.g., see the region shaded in purple in Figure \ref{fig:plp_cur}). 
As the PLP curves \cyc{based on different kernel sizes} \sunnycyc{typically show artifacts at different time positions,}
we find combining these PLP curves by element-wise multiplication a simple yet effective way of reducing the artifacts. We therefore define the combined PLP \sunnycyc{function $\Gamma^{\text{com}}$ by
\begin{equation} \label{eq:plp_comb}
\Gamma^{\text{com}}(n):=\Gamma^{1}(n)\cdot\Gamma^{3}(n)\cdot\Gamma^{5}(n)  
\end{equation} 
for $n\in [1:N]$.}

\subsection{PLP as Tempo-Related Condition}
\label{sec:plp_ref}

\begin{figure}
\centering
	\includegraphics[width = 0.9\columnwidth]{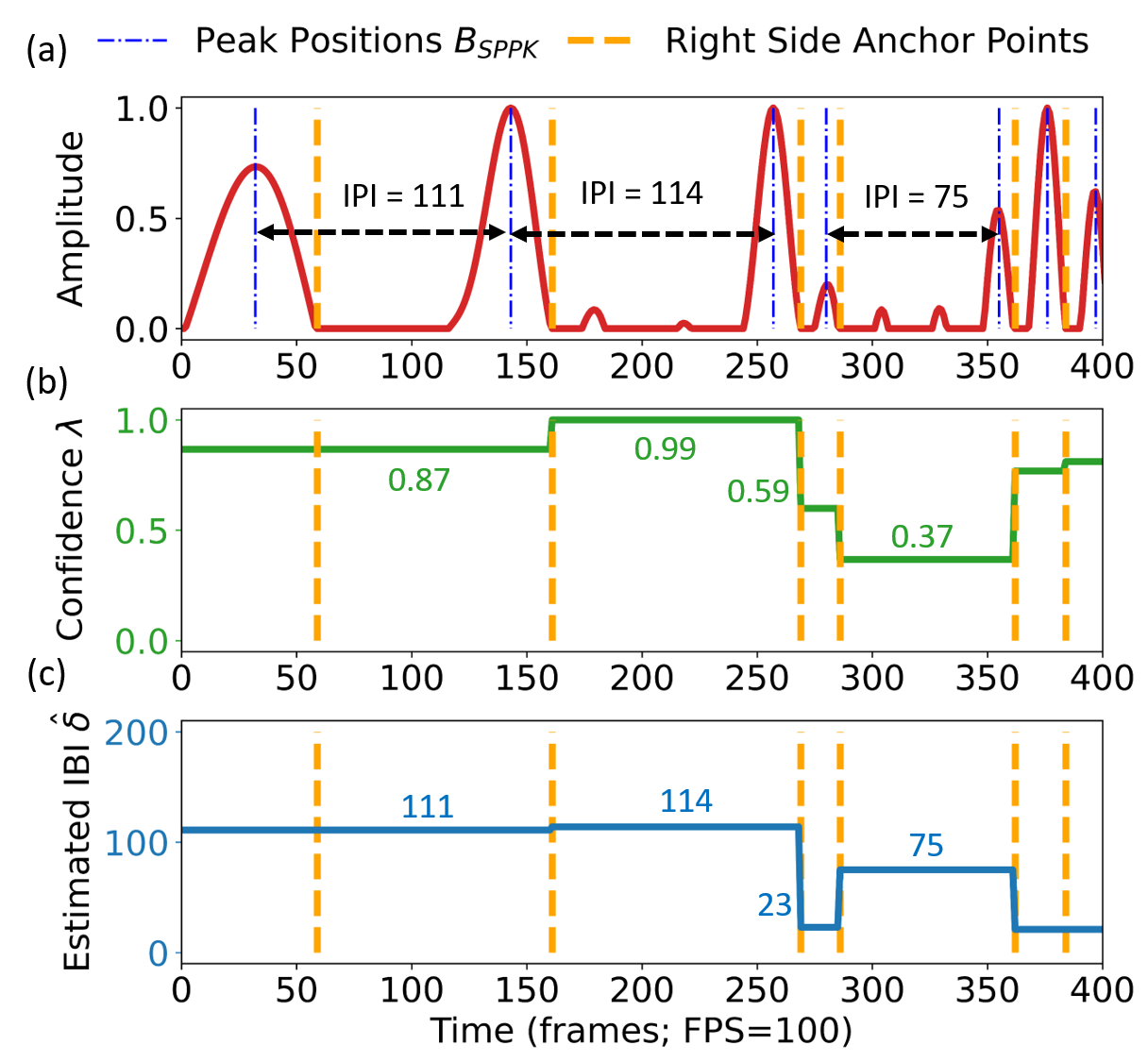}
	\caption{
 \sunnycyc{Conversion of a PLP function into piecewise constant functions of confidence $\lambda$ and estimated IBI $\hat{\delta}$. \textbf{(a)} \sunnycyc{Division of a PLP function into segments using right side anchor points of detected peaks.} \textbf{(b)} Confidence function \sunnycyc{derived from peak heights}. \textbf{(c)} Estimated IBI function $\hat{\delta}$ \sunnycyc{derived from inter-peak-intervals (IPIs)}. }
	}
	\label{fig:plpdp_ref}
\end{figure} 

PLP \sunnycyc{functions} yield \emph{peak positions} that are aligned with peaks in the input novelty function \sunnycyc{while the} \emph{peak heights} can be regarded as a measure of confidence.
\cyc{To obtain local tempo-related information, w}e propose the following procedure to convert a PLP \sunnycyc{function} into \sunnycyc{a piecewise constant function $\lambda$ expressing confidence and a piecewise constant function $\hat{\delta}$ encoding inter-beat-intervals (IBIs).} 
As exemplified in Figure \ref{fig:plpdp_ref}, 
we first apply a simple \emph{peak picking} function (SPPK)\footnote{\label{sppk}In our implementation, we use  \texttt{scipy.signal.find\_peaks} \cite{scipy} with parameters $\mathtt{height = 0.1}$, $\mathtt{distance =7}$, and $\mathtt{prominence = 0.1}$. The distance value of \sunnycyc{seven} frames is set \sunnycyc{to correspond} to the tolerance window size (i.e., 70 ms) for beat tracking evaluation, and the other two values are set \sunnycyc{to} 0.1 to ensure basic height and prominence of peaks. } to the PLP \sunnycyc{to obtain a list of}
peak positions $B_{\text{SPPK}}=(b_1, b_2, ..., b_K)$ (marked by blue lines in Figure \ref{fig:plpdp_ref}a).
We then divide the PLP into a number of segments at time \sunnycyc{instances} corresponding to the right side anchor points of these peaks (marked by vertical orange lines).\footnote{\cyc{While there might be other more complicated methods, we chose this simple heuristic that divides the PLP curve at the right side anchor points (i.e., positions where the curve reaches a low value after a peak).}}
\sunnycyc{For each such segment, we compute the inter-peak-interval (IPI) and set $\hat{\delta}(n)$ to this value for all frames $n$ within the segment, see Figure \ref{fig:plpdp_ref}c. Similarly, we define $\lambda(n)$ to be the average height of the segment's two PLP peaks.}
We show below \sunnycyc{how} the resulting two functions, confidence and estimated IBI, can be employed by a PPT for tracking the beats in expressive music.

\subsection{Combination of PLP and DP}
\label{sec:plp_dp_comb}
\begin{figure}
\centering
	\includegraphics[width = 0.9\columnwidth]{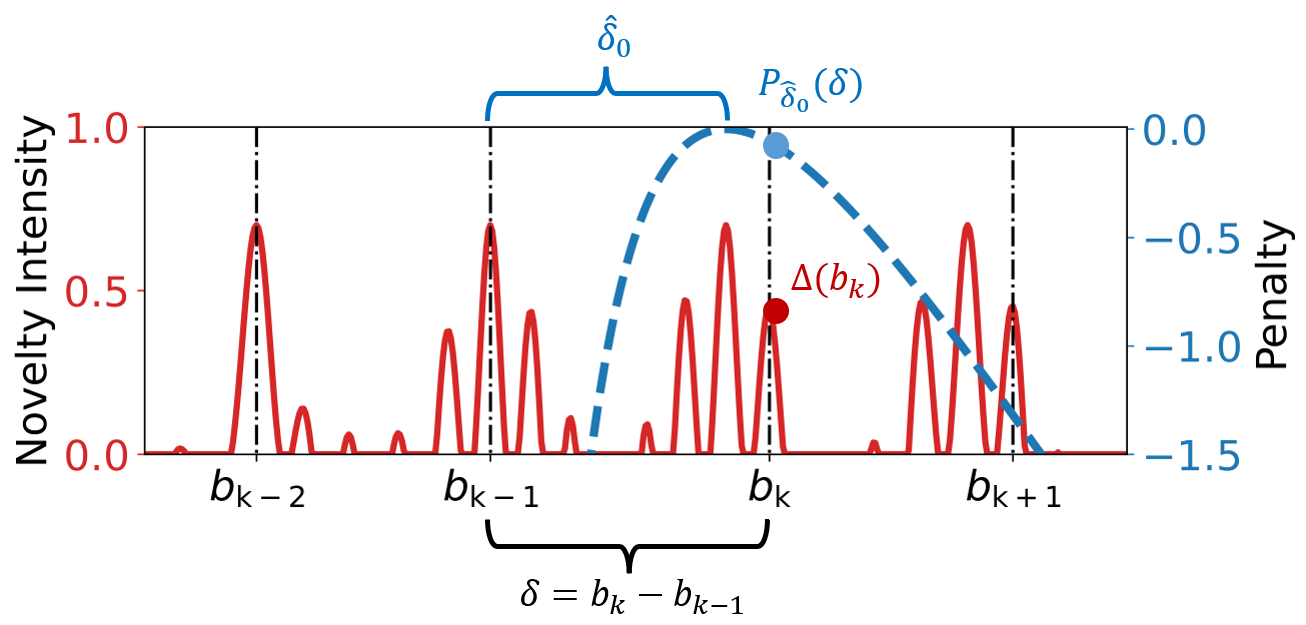}
	\caption{Illustration of the score \sunnycyc{function} $C(B)$ defined in Equation (\ref{eq:dp_score}), which jointly \sunnycyc{shows} the novelty function intensity $\Delta(b_k)$ and the tempo consistency penalty function $P_{\hat{\delta}_{0}}(\delta)$.
	}
	\label{fig:penalty}
\end{figure}

\sunnycyc{The DP-based beat tracker introduced in \cite{DP2007} aims at finding the optimal}
beat sequence $B^{\ast}$ that maximizes a score function $C$ 
\sunnycyc{balancing out}
novelty intensity and tempo consistency. Let $B=(b_1, b_2, ..., b_K)$ be a sequence of estimated beat positions in chronological order. The score $C$ \sunnycyc{function is defined by}:
\begin{equation} \label{eq:dp_score}
C(B) := \sum_{k=1}^{K}\Delta(b_k)+\lambda_0\sum_{k=2}^{K}P_{\hat{\delta}_0}(b_k-b_{k-1}) \,,
\end{equation}
where $\Delta$ denotes the novelty function, $\lambda_0 \in \mathbb{R}_{\geq 0}$ denotes a factor to balance the relative importance of the novelty function and the tempo consistency condition. \sunnycyc{Furthermore,} $P_{\hat{\delta}_0}:\mathbb{N}\rightarrow\mathbb{R}$ denotes \sunnycyc{a} penalty function for tempo consistency 
\sunnycyc{with respect to}
a preassigned IBI $\hat{\delta}_0\in \mathbb{N}$  \sunnycyc{defined by}
\begin{equation} \label{eq:dp_penalty}
P_{\hat{\delta}_0}(\delta) := -\left(\log_{2}\left(\frac{\delta}{\hat{\delta}_0}\right)\right)^2 \,
\end{equation}
\sunnycyc{for $\delta = b_k-b_{k-1}$. Note that $P_{\hat{\delta}_0}(\delta)$ is large for $\delta \approx {\hat{\delta}_0}$ and decreases for smaller or larger $\delta$ values. See Figure \ref{fig:penalty} for an illustration of $P_{\hat{\delta}_0}$ and the definition of the score function $C$.}

The original DP \cite{DP2007, libfmp2021},
\sunnycyc{\cite[Section 6.3.2]{fmp2021}}
takes a fixed IBI $\hat{\delta}_0$ and a fixed factor $\lambda_0$. 
In this paper, we propose a new DP-based PPT, named PLPDP, that takes \sunnycyc{the novelty function $\delta$} and the time-varying \sunnycyc{values} $\hat{\delta}(n)$  and $\lambda(n)$ introduced in Section \ref{sec:plp_ref} as input.
\sunnycyc{Compared to the original DP}, \cyc{we} replace $\hat{\delta}_0$ by $\hat{\delta}(n)$, and replace $\lambda_0$ by $\lambda(n)$. 
\sunny{In the DP \emph{forward} procedure, we compute at time frame $n$ an accumulated score $\textbf{D}(n)$ which depends on the accumulated scores of the previous frames $m\in [1:n-1]$, the corresponding tempo consistency penalties, and current novelty intensity: 
\sunnycyc{
\begin{align*}
& \textbf{D}(n) = \Delta(n) +\\ 
& \max\left\{0, \max_{m\in[1:n-1]}\left\{\textbf{D}(m)+\lambda(n)P_{\hat{\delta}(n)}(n-m)\right\}\right\}.
\end{align*}}
At the same time, we save in $\textbf{P}(n)$ the predecessor time position of the maximum. $\textbf{D}(n)$ and $\textbf{P}(n)$ are then used in the \emph{backward} procedure to derive the optimal beat sequence $B^{\ast}$}.  
Algorithm \ref{alg:alg1} shows the pseudo-code of PLPDP.\footnote{This algorithm is modified based on the DP algorithm in \cite{fmp2021} pp.343, Table 6.1. and \cite{fmpdp}.}

\begin{algorithm}[t]
\caption{PLPDP Beat Tracking}\label{alg:alg1}
\begin{algorithmic}
\STATE 
\STATE {\textsc{INPUT}}\\
\hspace{0.2cm} -- novelty (activation) function 
\sunnycyc{$\Delta:[1:N]\rightarrow [0,1]$}\\
\hspace{0.2cm} -- confidence \sunnycyc{$\lambda:
[1:N]\rightarrow [0,1]$ and}\\
\hspace{0.5cm} estimated IBI  $\hat{\delta}:[1:N]\rightarrow\mathbb{R}_{\geq 0}$ derived from PLP\\
\STATE {\textsc{OUTPUT}}\\
\hspace{0.2cm} -- optimal beat sequence $B^{\ast} = (b_{1}, b_{2}, ..., b_{K})$\\
\hspace{0.2cm} -- accumulated score $\textbf{D}(n)$\\
\hspace{0.2cm} -- predecessor information $\textbf{P}(n)$

\STATE {\textsc{PROCEDURE}}\\
\STATE {\textit{\textbf{Forward}}:}\\
\hspace{0.2cm}Initialize $\textbf{D}(0) = 0$ and $\textbf{P}(0) = 0.$\\
\hspace{0.2cm}Then compute in a loop for $n=1,..., N$:\\
\hspace{0.2cm}$\textbf{D}(n) = \Delta(n) +$\\ \hspace{0.5cm}$\max\left\{0, \max_{m\in[1:n-1]}\left\{\textbf{D}(m)+\lambda(n)P_{\hat{\delta}(n)}(n-m)\right\}\right\}$\\
\hspace{0.2cm}If $\textbf{D}(n) = \Delta(n)$ then set $ \textbf{P}(n)=0$, \\
\hspace{0.2cm}otherwise set \\
\hspace{0.5cm}$\textbf{P}(n) = \arg\max_{m\in[1:n-1]}\left\{\textbf{D}(m)+\lambda(n)P_{\hat{\delta}(n)}(n-m)\right\}$\\

\STATE {\textit{\textbf{Backward}}:}\\
\hspace{0.2cm}Set $k=1$ and $a_{k}=\arg\max_{n\in[0:N]}\textbf{D}(n)$\\
\hspace{0.2cm}Then repeat the following steps until $\textbf{P}(a_{k})=0$:\\
\hspace{0.5cm}Increase $k$ by one.\\
\hspace{0.5cm}Set $a_{k}=\textbf{P}(a_{k-1})$.\\

\hspace{0.2cm}If $a_{k}=0$, then set $K=0$ and \textbf{return}  $B^{\ast}=\emptyset$.\\
\hspace{0.2cm}Otherwise let $K=k$ and \\
\hspace{0.5cm}\textbf{return} $B^{\ast}=(a_{K}, a_{K-1},..., a_{1})$.

\end{algorithmic}
\label{alg1}
\end{algorithm}

We consider as the default case the combined PLP \sunnycyc{function $\Gamma^{\text{com}}$ as input} for PLPDP. To empirically justify the use of multiple kernels, we also consider in our experiments \cyc{an ablation study, where we use only \sunnycyc{the PLP function} $\Gamma^{\text{3}}$ with a single kernel, dubbed ``PLPDP-$\Gamma^{3}$.''} 

\section{Experiment Setup}
\label{sec:exp_setup}
\sunnycyc{In the following, we report on experiments to evaluate}
\sunnycyc{previous} state-of-the-art and \sunnycyc{our} proposed PPTs, see Section~\ref{sec:results}. \sunnycyc{In these experiments, we consider two datasets (Maz-5, ASAP), see Section \ref{sec:dataset_stat}. Furthermore, we consider activation functions computed from audio recordings (real use case) and obtained from GT annotations (synthetic scenario).}

\subsection{Statistics of Datasets}
\label{sec:dataset_stat}
Table \ref{tab:dataset} lists the two datasets employed in this study.
Maz-5 \cite{Grosche2010b} is a private collection of music composed of 301 audio recordings corresponding to five of the 49 different Chopin Mazurkas. \sunnycyc{These recordings were collected as part of the Mazurka project \cite{maz2010} and manually annotated (beat positions) by Sapp \cite{Sapp2008}}.
ASAP \cite{asap-dataset} is a public dataset newly released in 2020, consisting of 502 performances of Western classical piano music from 15 composers.\footnote{During the execution of this project, a dataset called ACPAS\cite{liu2021acpas} that combines ASAP and another 59 real recordings of classical piano performances was released. We \sunnycyc{have} the analysis and evaluation for those relatively small number of recordings not included in this work.}

Table \ref{tab:dataset} also shows the \emph{tempo stability rate} for the  datasets, calculated according to the approach of Schreiber \cite{Schreiber2020a}.
Specifically, we first convert all IBIs of a dataset into tempo values, and divide these tempo values by the average tempo of the corresponding track to derive normalized tempi. With the commonly adopted $\pm 4\%$ tolerance interval for quantifying whether the tempi of a recording \sunnycyc{are} stable or not \cite{gouyon2006}, we calculate the percentage of recordings in a dataset whose normalized tempi falls between the 0.96--1.04 interval.
Table \ref{tab:dataset} shows that the tempo stability rate of Maz-5 and ASAP is $13.1\%$ and $24.6\%$, respectively, suggesting that a large portion of the recordings in both datasets do not have a steady tempo.\footnote{In comparison, as reported \cite{Schreiber2020a}, the percentage of recordings with stable tempi reaches $90.9\%$ for the \emph{Ballroom} dataset \cite{gouyon2006, ballroom2}, a collection of dance music widely used in research on beat and downbeat tracking.}

\begin{table}[t]
\caption{Statistics of Datasets Used in the Experiments}
\centering
\begin{tabular}{c|c|c|c}
\toprule
\textbf{Dataset} & \textbf{\# tracks} & \textbf{Total duration} & \textbf{\% of stable tempi} \\
\midrule
\textbf{Maz-5}   & 301       & 12h 27m        & 13.1\%             \\
\textbf{ASAP}    & 502       & 41h 45m        & 24.6\%      \\\bottomrule      
\end{tabular}
\label{tab:dataset}
\end{table}

\subsection{Baseline/Proposed PPTs} 
\label{sec:ppt_settings}
We consider in our experiments four baselines \sunnycyc{PPTs}(SPPK, DP, mHMM, and mHMMT0) \sunnycyc{as well as our procedures,} PLPDP, and PLPDP-$\Gamma^{3}$. \sunnycyc{In the following, we describe these PPTs in more detail.} 

\subsubsection{Simple peak-picking (SPPK)} \cyc{This procedure applies} the \texttt{find\_peaks} function \sunnycyc{from the} \texttt{scipy} library \cite{scipy} to detect peak positions \sunnycyc{in the activation function. These peaks are taken as beats without }
imposing assumptions such as those regarding the beat periodicity consistency, and without any corrections of the input (e.g., novelty function enhancement)\footnotemark[3]. 

\subsubsection{DP+GT}
As mentioned in Section \ref{sec:plp_dp_comb}, DP \cyc{from \cite{DP2007}} requires a pre-assigned IBI $\hat{\delta}_0$ as global tempo information, which could be estimated by a global tempo detection \sunnycyc{approach} \cite{schreiber2020tempo}. 
However, as the focus here is to investigate the intrinsic limitations of DP, we use the ground-truth (GT) global tempo derived from the mean IBI of the reference beat positions.  \sunnycyc{Note that using GT tempo values gives} DP advantages over the other \sunnycyc{uninformed} PPTs \sunnycyc{thus avoiding} error propagation from imperfect global tempo estimation.
Moreover, we will show in our experiments that, even with access to some GT information, this DP+GT baseline does not perform well for Maz-5 and ASAP, \sunnycyc{due to its intrinsic limitations}. \sunnycyc{In our experiments, we use GT-informed}
DP following the settings of Grosche \emph{et al.} \cite{Grosche2011}\footnote{Source codes of DP can be found in \cite{libfmp2021}.}.

\subsubsection{mHMM and mHMMT0}
\label{sec:HMMtrans_model}
\sunnycyc{As another baseline approach, we employ the classic HMM-based PPT proposed by Krebs \emph{et al.} \cite{Krebs2015, Bock2016mm}\footnote{\sunnycyc{We adopted the official released \texttt{DBNBeatTrackingProcessor} of \texttt{madmom}, which does not consider the time signature of input tracks.}}. \sunnycyc{The main components of this approach are a}
state-space discretization and tempo transition model. 
Given an input novelty function $\Delta :[1:N]\rightarrow [0,1]$ as the observation sequence, the tempi to be considered are represented by a set

\begin{equation}
    \mathcal{A}:=\{\alpha_1, \alpha_2, ..., \alpha_I\}
\end{equation} of size $I\in \mathbb{N}$ consisting of distinct elements $\alpha_i$ for $i\in[1:I]$. The elements $\alpha_i$ are referred to as hidden tempo states determined by the discretization methods and a given tempo range. The tempo transition can be realized by a system that can be described at any time instance $n\in [1:N]$ as being in one of the tempo states $\dot{\Psi}_n \in \mathcal{A}$.

Kreb \emph{et al.}\cite{Krebs2015} proposed a tempo transition model that mostly stays in the same tempo and only allows tempo changes at beat positions. For a time instance $n$ that corresponds to a beat position, they empirically adopt the following exponential distribution function as the tempo change likelihood function:
\begin{equation} \label{eq:tmp_trans}
	f(\dot{\Psi}_n, \dot{\Psi}_{n-1}) = \exp\left(-\lambda_\mathrm{trans}\cdot\left| \frac{\dot{\Psi}_n}{\dot{\Psi}_{n-1}} -1 \right|\right) \,,
\end{equation}
where the tempo transition lambda $\lambda_\mathrm{trans}\in \mathbb{R}_{\geq0}$ determines the steepness of the distribution.\footnote{Note that the tempo discretization ($\dot{\Psi}$) of mHMM state-space is nonlinear and different from $\theta$ (in Section \ref{sec:plp}). Readers may refer to \cite{Krebs2015} for details. In this work, we set the tempo range of both mHMMs and PLPDP as 30--300\,BPM. 
I.e., $\theta \in [30:300]$, and \cy{(\texttt{min\_bpm}, \texttt{max\_bpm}) $=(30, 300)$ for \texttt{DBNBeatTrackingProcessor} of \texttt{madmom}.}} \sunnycyc{Intuitively speaking, using a large parameter $\lambda_{\mathrm{trans}}$ makes the model rigid, only allowing small tempo changes from beat to beat. Conversely, a small parameter $\lambda_{\mathrm{trans}}$ (close to zero) makes a transition to all possible tempi almost equally likely.}

}

By default, the tempo transition lambda $\lambda_\mathrm{trans}$ in \sunnycyc{Equation~\ref{eq:tmp_trans}} for mHMM is empirically \sunny{chosen as} $\lambda_\mathrm{trans}=100$ based on empirical observations from mainstream pop, rock, or dance music. We denote the default one as mHMM \sunnycyc{and refer to \cite{Krebs2015} for further details.} To allow for a much larger tempo variation, as encountered in classical music, we implement a variant of the mHMM with $\lambda_\mathrm{trans}=0$ and denote it as mHMMT0. \sunnycyc{To further reveal the capability and limitation of the HMM-based PPTs, experiments of grid search for $\lambda_\mathrm{trans}$} are reported and discussed in Section \ref{sec:gs}.

\subsubsection{PLPDP and PLPDP-$\Gamma^{3}$}
There are several options for implementing the PLPDP. For example, the hop size $h$, the kernel size $\kappa$, or \sunnycyc{design choices} for extracting tempo-related conditions from PLP, could all be optimized in some way (e.g., grid search). However, as our focus is on investigating the general behaviors of these local temporal-based methods rather than optimizing these methods for a specific type of dataset, we omit such optimization/fine-tuning processes. We empirically \cyc{set} $\kappa=1, 3, 5$ seconds for the combined PLP ($\Gamma^{\text{com}}$) as input of PLPDP, and $\kappa=3$ seconds for the PLP ($\Gamma^{3}$) \sunnycyc{as input of PLPDP-$\Gamma^{3}$ used in our ablation study. We set the tempo range to $[60:300]$ (given in BPM) for $\kappa=1$ and to $[30:300]$ (given in BPM) for $\kappa=3, 5$ to ensure that each PLP kernel can at least accommodate one completed sinusoidal wave for each given tempo.}


\subsection{\sunnycyc{Real Use Case}}
\label{sec:real_exp}
\sunnycyc{In the real use case, we take original audio recordings as the input}
and derive for each \sunnycyc{recording} the beat activation and downbeat activation (indicating probability of beat and downbeat at each frame) via \texttt{RNNDownBeatProcessor} of \texttt{madmom} \cite{Bock2016d, Bock2016mm}\cyc{, which is a DL-based approach}. We then take the maximum \sunnycyc{of the two activation functions} at each frame to derive a joint beat activation \cyc{function as input of the various PPTs}. 
\subsection{\sunnycyc{Synthetic Scenario}}
\label{sec:synth_exp}
\sunnycyc{Rather than computing activation functions from audio recording, we also consider a synthetic scenario where using idealized activation functions derived from ground-truth beat annotations. To this end, we}
transform the annotation into \sunnycyc{a pulse train of equal pulse magnitudes using} a frame rate of 100 FPS. This way, the input activation \sunnycyc{functions} \cyc{are} ``perfect'' as they are all \cyc{of} the maximum strength (\sunnycyc{set to} $1-\epsilon$) at beat positions and with minimum values (\sunnycyc{set to} $\epsilon$) at non-beat positions.\footnote{A small value $\epsilon = 10^{-6} $ is needed to prevent error warnings of the \texttt{madmom} API \cite{Bock2016mm} for our implementation of mHMM.} 
We expect the synthetic experiments \sunnycyc{based on synthetic activation functions} to reveal the sensitivity of the PPTs to tempo stability without the influence caused by \sunnycyc{errors in the} estimated activation functions.


\section{Experiment Results}
\label{sec:results}
\sunnycyc{In this section, we report on our experiment results for the real use case and synthetic scenario. We use a tolerance window of $\pm$ 70ms to calculate the recall (R), precision (P), and F-measure (F1) as the performance metrics.}
\subsection{Quantitative Result on Maz-5}
\sunnycyc{We start our discussion with the Maz-5 dataset. Table \ref{tab:res_maz} shows the beat tracking results for various settings.}

For the real activation case, we first look at the results for SPPK to get some insights regarding the properties of the \mbox{Maz-5} dataset and corresponding activation functions. As SPPK picks all activation peaks as beat positions, we can infer from the high precision value (\sunnycyc{$P=0.918$}) that most activation peaks correspond to beat positions and infer from the recall value (\sunnycyc{$R=0.754$}) that there are some missing peaks in Maz-5. 
\sunnycyc{Next, note that}
the baseline PPTs (e.g., DP and mHMM) \sunnycyc{can} hardly achieve \sunnycyc{an} F1 score higher than $0.6$, which is in contrast to their superior performance reported in references \cite{Bock2016d, Bock2020, Grosche2011} for music with steady tempo. From the lower recall values of DP and HMM-based methods \sunnycyc{(DP: $R=0.501$, mHMM: $R=0.393$, mHMMT0: $R=0.450$) \sunnycyc{compared to} SPPK ($R=0.754$)}, we can further see that their poor performance is mainly due to ignorance of activation peaks. Besides, from the precision values of \sunnycyc{DP ($P=0.475$) and mHMM ($P=0.753$) which are lower than SPPK ($P=0.918$)}, we can see that DP and mHMM insert beat estimations at positions without activation peaks based on their \sunnycyc{strict} tempo assumptions. 

On the other hand, PLPDP behaves differently \sunnycyc{than} other PPTs. From the remarkably high recall (PLPDP-$\Gamma^{3}$: $R=0.936$, PLPDP: $R=0.917$) \sunnycyc{compared to} SPPK ($R=0.754$), we see the effectiveness of ``local temporal expectations'' to compensate for the missing activation peaks at beat positions. On the contrary, from the lower precision values (\mbox{PLPDP-$\Gamma^{3}$}: $P=0.696$, PLPDP: $P=0.777$) \sunnycyc{compared to} SPPK ($P=0.918$), we see that PLPDP also make false-positive estimations based on its local temporal expectations. Overall, the above behavioral differences between PLPDP and existing PPTs lead to a \sunnycyc{substantial} performance gap of F1 score (PLPDP: $F1=0.838$ vs. mHMMT0: $F1=0.595$ and mHMM: $F1=0.499$), indicating the effectiveness of ``local temporal periodicity'' for Maz-5. Additionally, the high F1 score of SPPK ($F1=0.822$) implies that for an expressive music \sunnycyc{recording} with high-quality activation functions (e.g., with fewer number of non-beat activation peaks), SPPK may achieve a high F1-score. 

\sunnycyc{The results of the synthetic case provide further insights into the above observations.}
\sunnycyc{Note} that taking the perfect synthetic activation functions as input, one may expect all PPTs to achieve \cyc{an} F1 score of $1.0$. However, except for SPPK \sunnycyc{(the only PPT without any tempo-related assumptions or restrictions)}, none of the other PPTs could achieve so. From the imperfect recall and precision, we can see that the strong global tempo assumptions of DP and mHMM not only lead to \sunnycyc{discarding} activation peaks, but also \sunnycyc{introduce} false-positive beat predictions in regions without any activation peaks. Observing these inherent limitations,
\sunnycyc{the} poor performance \sunnycyc{of DP and HMM} in the real activation experiments may be less surprising.

It can also be seen from the higher recall and precision of PLPDP that the proposed method can adapt better to the local tempo variations of Maz-5 when the input activation is perfect. Accordingly, PLPDP is more flexible than DP and mHMM.\footnote{\sunny{One may argue that mHMMT0 outperforms PLPDP in this synthetic case of Maz-5. We would like to note that the main purpose of synthetic experiments is to investigate the limitations of the assumptions of each PPT. With the most flexible setting of lambda, mHMMT0 is indeed more flexible than PLPDP (though with difference $< 1.2\%$). However, this flexible setting also remarkably limits the performance of mHMMT0 in real (imperfect) activation cases. Besides, we note that when the optimal \sunnycyc{activation function} is achievable (e.g., the DL-based networks really learn the comprehensive idea of beats as humans), the best PPT is always SPPK (i.e., without any assumption).}} 
Moreover, the higher precision of PLPDP compared to PLPDP-$\Gamma^{3}$ suggests the effectiveness of a combined PLP function rather than a single-kernel PLP function.

\begin{table}[t]
\caption{Beat tracking result on the Maz-5 dataset. \sunnycyc{The} two best scores per metric \sunnycyc{are highlighted} in bold}
\setlength{\tabcolsep}{5pt}
\centering
\begin{tabular}{l|ccc|ccc}
\toprule
\multirow{2}{*}{\textbf{PPT}} & \multicolumn{3}{c|}{\textbf{Real activation}}                                                                 & \multicolumn{3}{c}{\textbf{Synthetic activation}}                                                            \\ \cline{2-7}
                              & \multicolumn{1}{c}{F1} & \multicolumn{1}{c}{Recall} & \multicolumn{1}{c|}{Precision} & \multicolumn{1}{c}{F1} & \multicolumn{1}{c}{Recall} & \multicolumn{1}{c}{Precision} \\ \midrule
SPPK                          & \textbf{0.822}                           & 0.754                          & \textbf{0.918}                          & \textbf{1.000}                           & \textbf{1.000}                          & \textbf{1.000}                          \\\midrule
DP+GT                         & 0.488                           & 0.501                          & 0.475                          & 0.799                           & 0.808                          & 0.791                          \\
mHMM                           & 0.499                           & 0.393                          & 0.753                          & 0.794                           & 0.872                          & 0.732                          \\
mHMMT0                         & 0.595                           & 0.450                          & \textbf{0.903}                          & \textbf{0.994}                           & 0.994                          & \textbf{0.995}                          \\\midrule
PLPDP-$\Gamma^{3}$                      & 0.791                           & \textbf{0.936}                          & 0.696                          & 0.862                           & \textbf{1.000}                          & 0.766                          \\
PLPDP                         & \textbf{0.838}                           & \textbf{0.917}                          & 0.777                          & 0.982                           & 0.996                          & 0.968                         
 \\
\bottomrule
\end{tabular}
\label{tab:res_maz}
\end{table}

\subsection{Quantitative Result on ASAP}

Table \ref{tab:res_asap} shows the beat tracking results for \sunnycyc{the} ASAP \sunnycyc{dataset}. 
From the \sunnycyc{substantial} performance change of SPPK, \sunnycyc{we can conclude that there are significant differences between the properties of ASAP and Maz-5.}
Both recall (ASAP: $R=0.419$, Maz-5: $R=0.754$) and precision (ASAP: $P=0.607$, Maz-5: $P=0.918$) values drop dramatically. \sunnycyc{This indicates} that, for ASAP, the \texttt{madmom} network \sunnycyc{fails} to generate activation peaks at some beat positions \sunnycyc{while generating} \sunnycyc{a large number of spurious} activation peaks. Such differences may be caused by the properties of datasets (e.g., \sunnycyc{ASAP may} contain more non-beat note events) or \sunnycyc{insufficient training of the DL-based \texttt{madmom} network (not adapted to ASAP).} 
These observations further explain the results that none of the PPTs can achieve an F1 score higher than 0.50, which is different from the result for Maz-5. We recall that the tempo stability of ASAP is higher than Maz-5, as shown in Table \ref{tab:dataset}. Therefore, the results for ASAP \sunnycyc{indicate that the} poor beat tracking performance may also be caused by the properties of activation functions.


\begin{table}[t]
\caption{Beat tracking result on the ASAP dataset. \sunnycyc{The} two best scores per metric \sunnycyc{are highlighted} in bold}
\setlength{\tabcolsep}{5pt}
\centering
\begin{tabular}{l|ccc|ccc}
\toprule
\multirow{2}{*}{\textbf{PPT}} & \multicolumn{3}{c|}{\textbf{Real activation}}                                                   & \multicolumn{3}{c}{\textbf{Synthetic activation}}                                              \\\cline{2-7}
                              & \multicolumn{1}{c}{F1} & \multicolumn{1}{c}{Recall} & \multicolumn{1}{c|}{Precision} & \multicolumn{1}{c}{F1} & \multicolumn{1}{c}{Recall} & \multicolumn{1}{c}{Precision} \\\midrule
SPPK                          & 0.380                  & 0.419                      & \textbf{0.607}                         & \textbf{1.000}                  & \textbf{0.999}                      & \textbf{1.000}                         \\\midrule
DP+GT                         & 0.450                  & 0.458                      & 0.443                         & 0.903                  & 0.913                      & 0.894                         \\
mHMM                           & 0.473                  & 0.540                      & 0.500                         & 0.911                  & 0.947                      & 0.886                         \\
mHMMT0                         & 0.374                  & 0.324                      & \textbf{0.556}                         & 0.982                  & 0.986                      & \textbf{0.981}                         \\\midrule
PLPDP-$\Gamma^{3}$                      & \textbf{0.488}                  & \textbf{0.732}                      & 0.404                         & 0.829                  & \textbf{0.997}                      & 0.750                         \\
PLPDP                         & \textbf{0.493}                  & \textbf{0.707}                      & 0.418                         & \textbf{0.982}                  & 0.995                      & 0.971                       
 \\
\bottomrule
\end{tabular}
\label{tab:res_asap}
\end{table}

Despite the above mentioned differences \sunnycyc{between the two} datasets, similar behavioral patterns of PLPDP can be observed. From the higher recall of PLPDP \cyc{compared to} the other PPTs, we can see again the effectiveness of the ``local temporal expectations'' to compensate for the missing peaks at beat positions of activation functions. However, from the lower precision of PLPDP in ASAP \cyc{compared to} Maz-5, \sunnycyc{one may deduce} that with an increasing number of activation peaks at non-beat positions, the performance of PLPDP \sunnycyc{and PLPDP-$\Gamma^{3}$ drop substantially}. 

Similarly, the results of the synthetic case \sunnycyc{for} ASAP \sunnycyc{support} the above observations. 
The \sunnycyc{obvious fact} that SPPK has perfect scores for all three metrics\footnote{\label{asap-ann} Its recall rate is not $1.0$, mostly due to annotation errors (i.e., reference beats that are too close to each other and excluded by SPPK's $\mathtt{distance}=7$ setting.) caused by the semi-automatic annotating process of ASAP \cite{asap-dataset}.} again reflects how PPTs are limited by their \sunnycyc{intrinsic tempo} assumptions.
Comparing the F1 scores 
between Maz-5 and ASAP in the synthetic case, we also see that DP and mHMM are sensitive to low tempo stability and perform much better as the tempo gets \sunnycyc{more stable} in ASAP. In contrast, PLPDP is less sensitive to tempo changes and performs similarly for both datasets.

\subsection{Qualitative Results}
\label{sec:qualitative_res}

\begin{figure*}[t]
\centering
	\includegraphics[width = 2\columnwidth]{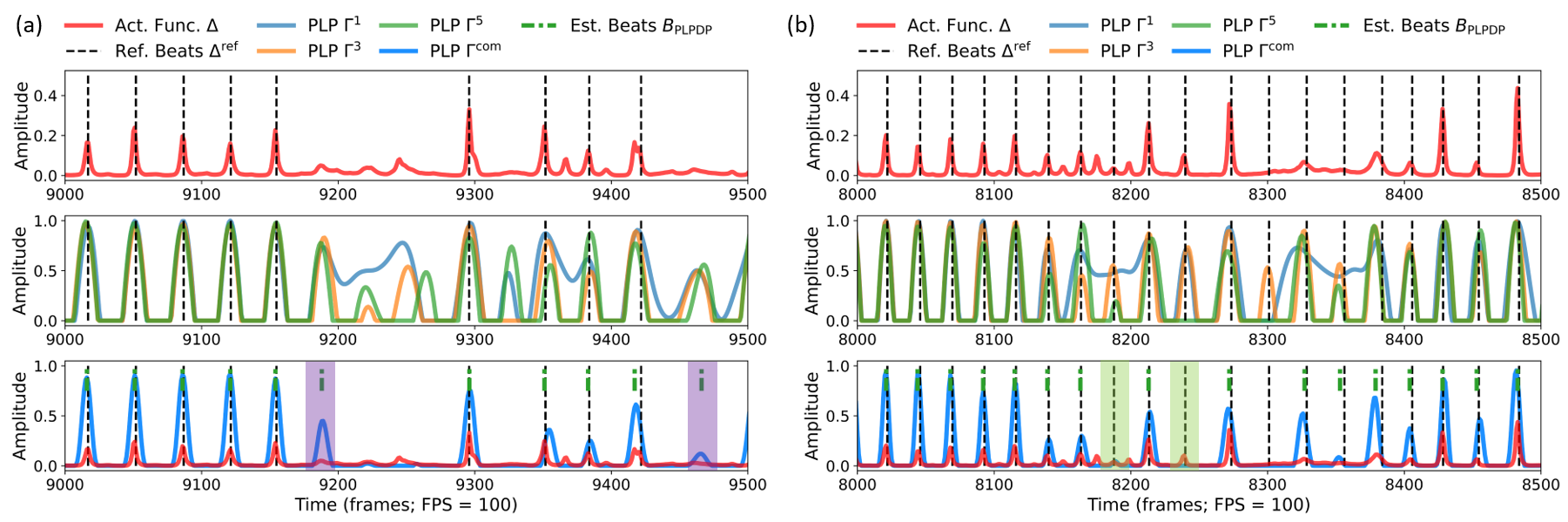}
	\caption{
	The beat tracking result of the proposed PLPDP-based PPT for the same two recordings used in Figure \ref{fig:teaser}. 
 \textbf{(a)} Maz-5 example. \textbf{(b)} ASAP example.
 \textbf{Top:} \texttt{madmom} activation function and reference beats. \textbf{Middle:} PLP functions with $\kappa=1, 3, 5$. \textbf{Bottom:} Combined PLP function and estimated beats of PLPDP. \sunnycyc{Purple-shaded regions} highlight the false-positive estimations. \sunnycyc{Green-shaded regions} highlight false-negative estimations. Best viewed in color.  
	}
	\label{fig:qualitative}
\end{figure*} 

Figure \ref{fig:qualitative} shows the beat tracking result of PLPDP for two real activation functions (continuing the examples from Figure \ref{fig:teaser}).
The reference beats and activation functions (Figure \ref{fig:qualitative}\sunnycyc{, top row}) illustrate the observations mentioned \sunnycyc{before}. For example, \sunnycyc{the} Maz-5 \sunnycyc{recording} has \sunnycyc{more} tempo variations and reveals fewer activation peaks at non-beat positions. On the other hand, we see several weak or missing activation peaks at beat positions of \sunnycyc{the ASAP example}. The PLP functions for $\kappa = 1, 3, 5$ (Figure \ref{fig:qualitative}\sunnycyc{, middle row}) further reveal the different temporal expectations at regions with tempo changes.
The combined PLP function and estimated beats of PLPDP (Figure \ref{fig:qualitative}\sunnycyc{, bottom row}) reveal the advantages and limitations of the local temporal expectations. Specifically, PLPDP nicely adapts \sunnycyc{for both examples} to the local tempo variations based on its local temporal expectations. However, \sunnycyc{these expectations may} also cause false-positive errors if there are activation peaks at non-beat positions that match the locally detected periodicity, as highlighted by the \sunnycyc{the purple-shaded regions}. From false-negative errors (\sunnycyc{green-shaded regions}), we see that PLPDP \sunnycyc{may} also ignore activation peaks at beat positions based on its local temporal expectations. However, as long as most of \sunnycyc{the} activation peaks are stronger at beat positions, PLPDP introduces much less such false-negative errors than mHMM.

\begin{figure*}[t]
\centering
	\includegraphics[width = 1.32\columnwidth]{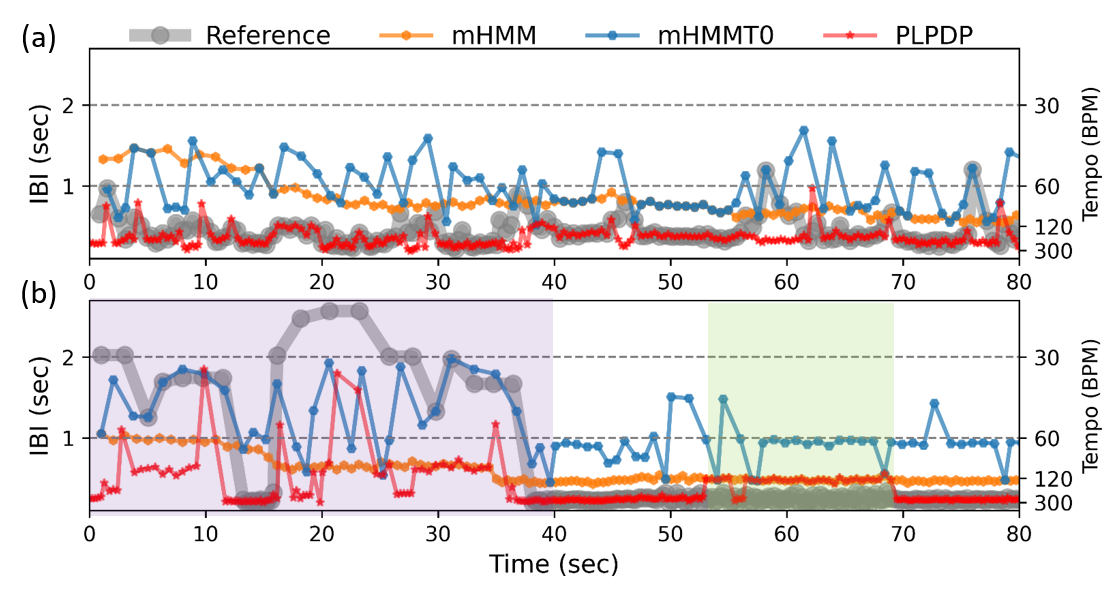}
	\caption{
	The IBI progression of reference beats (grey) and estimated beats (blue: mHMM, orange: mHMMT0, red: PLPDP) for \sunnycyc{the recordings of the excerpts used} in Figure~\ref{fig:teaser}. \sunnycyc{\textbf{(a)} Maz-5 recording. \textbf{(b)} ASAP recording.}
 \sunnycyc{Purple-shaded region} highlights a region with slow unstable tempi. \sunnycyc{Green-shaded region} highlights a region with fast stable tempi which PPTs fail to follow. 
	}
	\label{fig:IBIprog}
\end{figure*}

Figure \ref{fig:IBIprog} demonstrates the longer-term behavior of the PPTs for \sunnycyc{the recordings of the two} examples via plotting the inter-beat-interval (IBI) progression. Specifically, for each sequence of beat positions (e.g., reference beats or \sunnycyc{PPT-based} estimated beats), we \sunnycyc{include} the beat positions $b_i$ (in \sunnycyc{horizontal axis given in} seconds) and its corresponding IBI (i.e., \sunnycyc{the value $b_{i+1}-b_i$ plotted on vertical axis}) to see the reference/estimated IBI progression within \sunnycyc{the recordings}.
We can see from \sunnycyc{the} reference (grey) curve that while the \sunnycyc{Maz-5 example} (\sunnycyc{Figure \ref{fig:IBIprog}a}) \sunnycyc{reveals} continuous tempo changes at faster tempi (i.e., 100--300 BPM\sunnycyc{, corresponding to IBIs between 0.2--0.6 seconds.}), the ASAP \sunnycyc{example} (\sunnycyc{Figure \ref{fig:IBIprog}b}) has both \sunnycyc{an} slow unstable region (shaded in purple) and a region \sunnycyc{with faster stable tempi (shaded in green)}. For both pieces, any global assumption without explicitly \sunnycyc{considering} the local musical contents is not likely to work well. Explicitly, as the tempo transition function of mHMMs are set in a global manner, both mHMMs are not able to align with the reference IBI progression like PLPDP can do.

For more qualitative results, we refer \sunnycyc{the} readers to the project web page\sunnycyc{\footnotemark[1]}.

\subsection{Grid Search of mHMM Tempo Transition $\lambda_\mathrm{trans}$}
\label{sec:gs}

\begin{figure}[t]
\centering
	\includegraphics[width = 0.96\columnwidth]{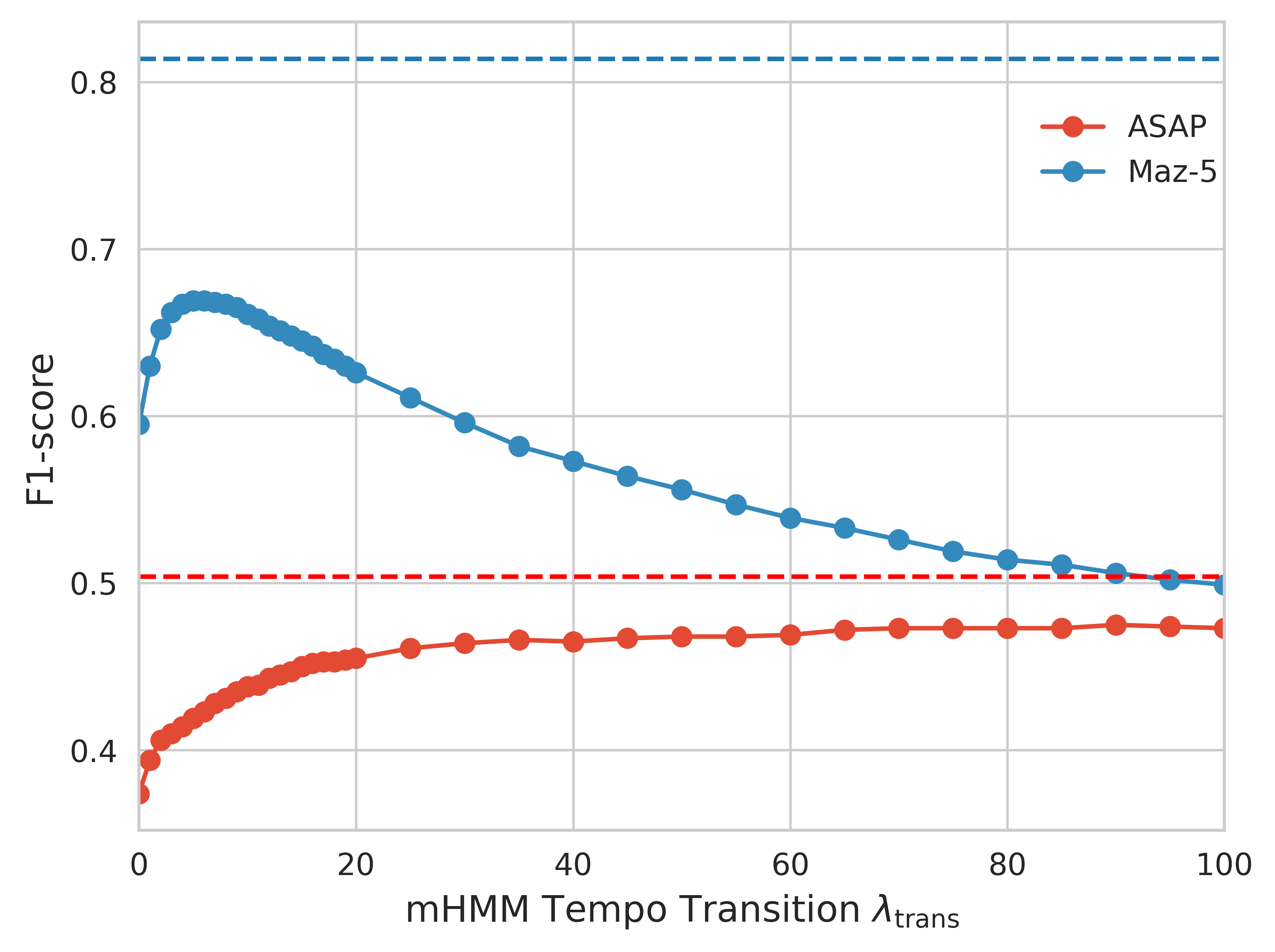}
	\caption{
	Grid search of mHMM tempo transition $\lambda_\mathrm{trans}$ from 0--100 for real activation experiments. A step size of one is adopted for $0 \leq \lambda_\mathrm{trans} \leq 20$, and five otherwise. Results of PLPDP are indicated as horizontal dashed lines for comparison.
	}
	\label{fig:gs_real}
\end{figure} 
\sunnycyc{The} above experiments have already demonstrated the \sunnycyc{conceptual} difference between HMM-based PPTs and the proposed PLPDP approach. \sunnycyc{We now present an additional grid-search experiment with regard to the mHMM tempo transition parameter $\lambda_{\mathrm{trans}}$ to provide some additional insights.}
Figure~\ref{fig:gs_real} shows the results for real the activation case. \sunnycyc{While the mHMM approache} with lambda $\lambda_\mathrm{trans}$ \sunnycyc{varying from 1 to 25} performs better than mHMMT0 for Maz-5, the best mHMM still performs worse than PLPDP (shown as horizontal dashed lines) for both datasets. Furthermore, though both datasets \sunnycyc{consist of expressive classical music}, the \sunnycyc{best performing} $\lambda_\mathrm{trans}$ are different (i.e., Maz-5: $\lambda_\mathrm{trans}=5$, ASAP: $\lambda_\mathrm{trans} = 90$).
\sunnycyc{One may therefore deduce that adjusting the parameter $\lambda_{\mathrm{trans}}$ for each recording and even to local sections within a recording may be essential to improve the performance of the mHMM approach.}

\section{Conclusion and Future Work} 
\label{sec:conclusions}
In this paper, we have \sunnycyc{made} contributions towards \sunnycyc{improving and better understanding} beat tracking for expressive classical music. First, we introduced a new local temporal expectation-based post processing tracking (PPT) method. Second, we conducted experiments to investigate the performance upper bounds of the considered PPTs. Third, we presented a comprehensive evaluation and analysis of beat tracking approaches for expressive classical music.
The proposed PLPDP method provides a way to incorporate local tempo-related information into a beat tracking system. \sunnycyc{Considering} local periodicity consistency, our method \sunnycyc{differs from} existing PPTs that rely on globally determined tempo transition assumptions. Moreover, \sunnycyc{our} synthetic experiments demonstrate new ways to investigate and explore the strength and limitations of  PPTs. \sunnycyc{Overall}, we hope this work provides a new direction towards \sunnycyc{improving} beat tracking for expressive classical music.

From the real activation experiments on ASAP, we  see that the PPTs still have \sunnycyc{a large margin for improvements.}
Among the factors that influence the performance of beat tracking, the missing peaks of activation at beat positions could be potentially reduced by adding more data of classical music for training the feature-learning networks. \sunnycyc{In particular, this may lead to substantial improvements of activation functions that can better account for the various properties of note onsets as occurring in classical music.}

Adding training data alone, however, might not help the DL-based networks to produce less (false-positive) non-beat activation peaks, which also constitute a large part of the beat tracking errors \sunnycyc{in particular} for ASAP. Instead of relying completely on onset-related information, we conjecture that approaches that consider hierarchical cues, e.g., frequency domain information as pitch, melody, or longer-term structure-related information, might prove useful. 
Besides, these \sunnycyc{hierarchically arranged musical and acoustic cues}
may also be helpful for future models to adaptively adjust PLP kernel sizes. 

From our empirical observations of the behavior of all the considered PPTs, we find that \sunnycyc{these procedures often} switch between different metric-levels (e.g., half, third, double or triple tempo of reference beats) while tracking the beats of a recording.
Such a ``metric-level switching'' behavior, however, cannot be reflected by the current evaluation metrics. \sunny{We have recently proposed an analysis method \cite{chiu2022acr} \sunnycyc{for gaining a better understanding of such issues}. More results and discussions can be found in our GitHub repository.}

Lastly, 
as existing datasets of multi-instrument classical music (e.g., RWC-Classical \cite{rwcdatabase}) are relatively small, we consider only Western classical piano music in \sunnycyc{our} experiments. 
Compared to piano music, there might be more onsets at non-beat positions in multi-instrument classical music, and the activation intensity at beat positions may be weaker due to soft onsets. 
To assess the performance of PLPDP for expressive classical music in general, future work is needed to consider datasets beyond piano music.

\bibliographystyle{IEEEtran}
\bibliography{taslp2022}

\begin{IEEEbiographynophoto}{Ching-Yu Chiu}
received the Ph.D degree (2023) from Graduate Program of Multimedia Systems and Intelligent Computing, National Cheng Kung University and Academia Sinica, Taiwan. She is currently a postdoctoral researcher at International Audio Laboratories Erlangen, a joint institute of the Friedrich-Alexander-Universit\"at Erlangen-N\"urnberg (FAU) and the Fraunhofer Institute for Integrated Circuits IIS. Her research interests include music information retrieval, signal processing, and machine learning.
\end{IEEEbiographynophoto}

\begin{IEEEbiographynophoto}{Meinard M\"uller}
received the Diploma degree (1997) in mathematics and the Ph.D. degree (2001) in computer science from the University of Bonn, Germany. After his postdoctoral studies (2001-2003) in Japan and his habilitation (2003-2007) in multimedia retrieval in Bonn, he worked as a senior researcher at Saarland University and the Max-Planck Institut für Informatik (2007-2012). Since 2012, he has held a professorship for Semantic Audio Signal Processing at the International Audio Laboratories Erlangen, a joint institute of the Friedrich-Alexander-Universit\"at Erlangen-N\"urnberg (FAU) and the Fraunhofer Institute for Integrated Circuits IIS. His recent research interests include music processing, music information retrieval, audio signal processing, and motion processing. He was a member of the IEEE Audio and Acoustic Signal Processing Technical Committee (2010-2015), a member of the Senior Editorial Board of the IEEE Signal Processing Magazine (2018-2022), and a member of the Board of Directors, International Society for Music Information Retrieval (2009-2021, being its president in 2020/2021). In 2020, he was elevated to IEEE Fellow for contributions to music signal processing. 
\end{IEEEbiographynophoto}


\begin{IEEEbiographynophoto}{Matthew E. P. Davies}
received the B.Eng. degree in computer systems with electronics from King's College London, U.K., in 2001, and the Ph.D. degree in electronic engineering from Queen Mary University of London (QMUL), London, U.K., in 2007. From 2007 until 2011, he was a Postdoctoral Researcher with the Centre for Digital Music, QMUL. In 2013, he worked in the Media Interaction Group, National Institute of Advanced Industrial Science and Technology. From 2014 to 2019, he coordinated the Sound and Music Computing Group, INESC TEC, and is currently a Researcher with the Centre for Informatics and Systems of the University of Coimbra. His main research interests include music information retrieval, evaluation methodology, and creative music systems.
\end{IEEEbiographynophoto}

\begin{IEEEbiographynophoto}{Alvin Wen-Yu Su}
(M’97) received the B.S. degree in control engineering from National Chiao-Tung University, Hsinchu, Taiwan, in 1986, and the M.S. and Ph.D. degrees in electrical engineering from Polytechnic University, Brooklyn, NY, USA, in 1990 and 1993, respectively. From 1993 to 1994, he was with the Center for Computer Research in Music and Acoustics, Stanford University, Stanford, CA, USA. Currently, he is a Professor of the Department of Computer Science and Information Engineering, National Cheng-Kung University, Tainan, Taiwan. His research interests cover the areas of physical modeling of acoustic musical instruments, data compression, audio/image/video signal processing, and VLSI.
\end{IEEEbiographynophoto}

\begin{IEEEbiographynophoto}{Yi-Hsuan Yang} (M’11-SM’17) received the Ph.D. degree in Communication Engineering from National Taiwan University. Since 2023, he has been with the College of Electrical Engineering and Computer Science, National Taiwan University, where he is a Full Professor. Prior to that, he was the Chief Music Scientist at an industrial lab called Taiwan AI Labs from 2019 to 2023, and an Associate/Assistant Research Fellow of the Research Center for IT Innovation, Academia Sinica, from 2011 to 2023. His research interests include automatic music generation, music information retrieval, and machine learning. 
He served as an Associate Editor for the \emph{IEEE Transactions on Affective Computing} and the \emph{IEEE Transactions on Multimedia}, both from 2016 to 2019. Dr. Yang is a senior member of the IEEE.
\end{IEEEbiographynophoto}

\end{document}